# The planar instability during unidirectional freezing of a polymer solution: Diffusion-controlled or not?

*Tongxin Zhang, Zhijun Wang\*, Lilin Wang\*, Junjie Li, Jincheng Wang*

*State Key Laboratory of Solidification Processing, Northwestern Polytechnical University, Xi'an 710072, China*

**Abstract:** Freezing of polymer solutions has been extensively investigated from many aspects, especially the complex pattern formation. The cellular/dendritic microstructures occurred in freezing of polymer solutions are usually believed to belong to the type of diffusion-induced Mullins–Sekerka (M-S) instability. However, the presence of macromolecule as an impurity in water is significantly different from that of small ions. The quantitative investigation on transient process of directional freezing of polymer solutions remains lack due to some challenges. For the first time, we in-situ observed the planar instability behaviors during unidirectional freezing of a polymer solution together with a typical ionic solution with manipulated ice orientation. It is found that unlike the ionic solution exhibiting typical diffusion-controlled planar interface instability, the solute recoil of a polymer solution deviated severely from the predictions of W-L model. Meanwhile, the real-time observation shows that the polymer solution exhibits a global instability mode instead of a local instability mode. These results reveal the complex physics behind freezing of a polymer solution, and are believed to promote relevant investigations in terms of the theoretical approach to describing the freezing behavior of a polymer solution.

## 1. Introduction

Pattern formation via water freezing has always been an attractive but sophisticated example in crystal growth regime [1]. In contrast, morphogenesis mechanism in freezing of polymer solutions has received far less attention outside the cryobiology community [2-7]. Yet, solidification of water is the dominant forming process for solutions with soluble macromolecules such as polyvinyl alcohol (PVA) [8], polyethylene glycol (PEG)[9], collagen [10] and chitosan [11, 12]. When a polymer is dissolved in water, swelling gradually occurs and the polymer becomes a solute with specific steric configuration such as coil and globule, depending on the solubility of the polymer in its solvent. Unlike the freezing of ionic solution with a relatively simple nature, the mechanism of pattern formation via freezing of polymer solution featuring hierarchical porous microstructure is reported to be the combined result of at least three conflicting processes: freezing of the solvent, phase separation of polymer solution, and crystallization of polymer [13]. Besides, the unique properties of polymer solution which arise from the nonlinear dependence of mass transport on polymer concentration [14-18], interface attachment processes [4, 19-21] and its other chemical properties have made freezing of polymer solution a peculiar subject in relevant pattern formation. Although some previous investigations provided useful information on freezing of polymer solution, the physics of its solid/liquid (S/L) interface instability, as a basic but intriguing question, is still unclear. The Mullins-Sekerka (M-S) instability is usually borrowed to explain the origin of pattern formation in a polymer solution [8, 10, 22, 23] and diffusion-controlled instability is usually believed to be the main mechanism of planar instability of a polymer solution. In fact, it is of interest to utilize the relevant solidification model to polymer solution as it provides a rather special test of the applicability of some solidification theories. The Warren-Langer (W-L) model has been proposed to describe planar interface movement in unidirectional solidification and has been verified both experimentally [24] and computationally via phase field simulation [25]. Hence it is natural to consider whether the W-L model can reproduce quantitively the S/L interface movement near planar instability of a polymer solution.

---

\* Corresponding author. zhjwang@nwpu.edu.cn
\* Corresponding author. W-Lilin@nwpu.edu.cn

Apart from the theoretical approach to answer the question, experimental measurement is also needed in terms of the transient process of planar instability for a polymer solution. Up to now, the transient freezing process in relation with S/L interface position at the early stage of planar instability of polymer solution has never been quantitatively presented in previous investigations due to some great challenges. The first challenge is the preparation of a single ice crystal with well-controlled orientation since orientation effect is also crucial in the resulting solidification pattern [26-28]. The second challenge is the quantitative investigation of S/L interface morphology based on dynamical information on S/L interface movement. A horizontal

Bridgeman solidification platform is often utilized as a powerful tool to reveal in-situ details of morphogenesis in freezing of ionic solutions [29] or colloidal suspensions [30]. In fact, a good knowledge of S/L interface movement of a sample solution during directional freezing is vital to the relevant analysis of directional freezing experiments. The shift from the initial interface position (static) to its steady state (in motion at constant pulling velocity) is termed as "interface recoil" [31]. It is reported that the dynamic recoil of S/L interface position in the dynamic transient stage is caused by three different factors [31]: **a**. the "instrumental recoil" that results from thermal exchanges caused by pulling; **b**. the "solute recoil" that corresponds to depression of the interface equilibrium temperature associated with the formation of solute boundary layer; **c**. an isotherm shift due to release of latent heat. Factor **b** is expected to play a role in interface instability since it is a direct indicator of solute redistribution in the vicinity of S/L interface, whereas factor **a** and factor **c** that contribute to thermal lag need to be eliminated for the analysis via W-L model.

In this paper, we have overcome the thermal lag by a design of side-by-side samples and precisely obtained the interface recoil of PVA solution and NaCl solution by solute redistribution only. Then, we precisely in-situ measured the transient movement and morphological changes of advancing solid/liquid (S/L) interfaces of PVA and NaCl solutions during their planar instability processes on a customized horizontal Bridgeman freezing stage. The quantitative experimental data of the transient S/L interface position were obtained via differential visualization (DV) method [32, 33] and compared to Warren-Langer (W-L) model [34]. The quantitative results in this paper were suggested to shed new light on understanding the different natures of pattern formation during directional freezing of a polymer solution and an ionic solution.

## 2. Experimental setup

Ultrapure water (18.25 MΩ) was chosen as the solvent for the preparations of solutions and the pure solvent in DV method. NaCl (AR, 99.5 %) and two types of PVA solutions (PVA103 (Mw = 31000, 96.8%-97.6% hydrolyzed) and PVA203 (Mw =31000, 86.5%-89% hydrolyzed) were applied as the solutes. NaCl solutions were prepared at room temperature by dissolving NaCl powder in ultrapure water; PVA solutions were prepared at 80°C by dissolving PVA powder in ultrapure water via magnetic stirring for one hour. Prior to freezing experiments, both solvent for DV method and solutions were degassed under vacuum condition for half an hour to prevent possible occurrence of air bubbles near the S/L interface during freezing. Rectangular glass capillary with an inner space dimension of 0.05 × 1 mm$^2$ (VitroCom brand) was adopted as sample cell for unidirectional freezing.

The experiments were performed in a unidirectional freezing manner with ice crystal grown in thin rectangular capillary glued on a silica glass sheet. In our experimental apparatus, the thermal gradient was realized by hot and cold zones separated by a gap of 2.5 mm. Both heating and cooling zones of copper blocks (120 × 100 × 10 mm$^3$) were connected to ethanol thermostat. The temperature of ethanol thermostat can change from −20.0 °C to 50.0 °C discretionarily by a temperature controller. The area where unidirectional freezing occurred was thermally insulated by a surrounding double-layer PMMA shell with wall thickness of 8 mm and an interlayer of air of same thickness between PMMA walls. **Figure 1** shows crystal orientation of ice and the schematic diagram of horizontal unidirectional freezing stage with measurement of S/L interface position of aqueous solutions. DV method was applied to eliminate the thermal recoil which originated from each capillary via a side-by-side design of two parallel freezing samples (pure water and solutions in thin rectangular glass capillaries) glued on a silica glass sheet with a cross section of 20 x 0.05 mm$^2$ (see the sketch map of top view of the sample in **Fig. 1 (b)**). It can be seen in **Fig. 1 (b)** that the cross section of silica glass sheet (20 x 0.05 mm$^2$) is 200 times as large as that of each capillary (inner dimension of 0.05 x 1 mm$^2$), and thermal conduction during freezing is thus mainly achieved by silica glass sheet instead of capillary, allowing a linear thermal gradient to be established on both sides of the S/L interface. And the thermal recoil which originated from freezing in each capillary is, therefore, physically equivalent to the thermal recoil of glass sheet itself (i.e. $\Delta Z_3 = \Delta Z_{sheet}$ as shown in **Fig. 1 (b)**). It should be noted that, prior to each freezing experiments, the directional solidification of single ice crystal into liquid was initiated by means of a step-by-step methodology as reported elsewhere [35-37]. The single ice crystal whose orientation was specially manipulated as the edge plane was grown in capillary in pure water and PVA solution and NaCl solution under an imposed thermal gradient $G$ (see the relation of direction $\vec{V}_p \parallel \vec{G} \parallel \{0001\}$ on the arrow in the upper part of **Fig. 1 (a)**). The translational pulling velocity was provided by a linear ball-screw driven stage. The rotation of the ball screw was actuated by a DC servo-drives motor through a decelerator to provide continuous movement of freezing sample. The whole pulling equipment was fixed on a large cast-iron experiment table to eliminate exotic mechanical vibration. The motor was fixed on micro-positioners in order to align the screw on the track accurately. The S/L interface position and microstructure evolution of the S/L interface was recorded by a CCD camera with 2580 × 1944 sensitive elements on a time-lapse video recorder. The local imposed thermal gradient for each run of directional freezing was precisely measured by a thermalcouple glued on the used freezing samples in translational motion of roughly the same velocity. The thermalcouple was connected with $7_{1/2}$ Digit Nano Volt/Micro Ohm Meter (34420A, Agilent brand) with sampling rate of one point per second. In addition, the ice crystal orientation was simultaneously detected through a pair of polarizers to guarantee that the crystal orientation remained unchanged during directional freezing experiments.

Two groups (**Group A/B**) of experiments were carried out for different purposes of the paper. **Group A** belonged to quantitative investigation on solute recoil via DV method, which addressed the transient interface recoil by solute redistribution during planar instability of NaCl solutions (five initial concentrations of 0.05 M, 0.1 M, 0.2 M, 0.3 M and 0.6 M) and PVA solutions (three initial concentrations of 0.25 wt.%, 5 wt.% and 10 wt.% and two degree of hydrolysis 96.8%-97.6% hydrolyzed and 86.5%-89% hydrolyzed) under pulling velocities approximately ranging from 3.86 um/s to 5.77 um/s. The pulling velocities in **Group A** were intentionally kept very low (less than 10 um/s) so that the translational motion of freezing samples can result in only a minor influence on the imposed thermal gradient [32]. The side-by-side sample in **Group A** consisted of one capillary with PVA solutions of various initial concentrations and degrees of hydrolysis and the other with ultrapure water as pure solvent for usage of DV method. The NaCl solutions with varying concentrations were also investigated in the same way. W-L model [34] was later applied so as to reproduce the variation of the S/L interface position of both NaCl and PVA solutions in **Group A**. It should be noted that the effect of thermal lag was surmounted by DV method so that solute recoil of any sample solution in motion (i.e. $\Delta Z_2$ in **Fig. 1 (a)**) can be measured for quantitative analysis with W-L model. **Group B** belonged to an in-situ comparison investigation on the different planar instability of PVA and NaCl solutions under the same pulling velocities and thermal gradient, which consisted of one capillary with 0.5 wt.% NaCl solution and the other with 5wt.% PVA solution. Four pulling velocities approximately ranging from 5.65 um/s to 40.16 um/s were used in **Group B**. It should be noted that, since this paper deals with the planar instability of ionic and polymer solutions, each freezing sample were kept static for a time interval of more than 30 minutes to make the S/L interface planar before directional freezing under every used pulling velocity.

### 3. Representation of the diffusion-controlled W-L model

For the reading consistency here, we represent the derivation of Warren-Langer (W-L) model [34] which can deal with the diffusion-controlled transient stage along with the build-up of solute boundary layer. Derivation of W-L model is started with the transient diffusion equation in a moving frame (directional solidification at pulling velocity $V_p$) in one dimension. The freezing system to be described by W-L model is an aqueous solution. The one-dimensional coordinate is shown in Fig. **1 (a)**, where $Z$ is the distance coordinate in the direction of pulling velocity $V_p$ whose original point $O$ is chosen as the position of the S/L interface of pure solvent. The governing solute diffusion equation Eq.1 and the corresponding boundary conditions Eq.2-4 for the freezing system are given below.

in the liquid phase: $\dfrac{\partial C_L}{\partial t} = D_L \dfrac{\partial^2 C_L}{\partial Z^2} + V_p \cdot \dfrac{\partial C_L}{\partial Z}$ (Eq. 1)

temperature distribution ahead of the S/L interface with a distance $Z$

$$T = T_M + G \cdot Z \tag{Eq. 2}$$

local equilibrium condition at the S/L interface

$$C_I(Z_I, t) = -\frac{G}{m_L} Z_I(t) \tag{Eq. 3}$$

solute mass balance at the S/L interface

$$\frac{\partial C_L}{\partial Z} = -\frac{V_I}{D_L}(1-k)C_I(Z_I, t) \tag{Eq. 4}$$

where $C_L$ is the solute concentration in the liquid phase beyond the S/L interface with a distance of $Z$, $Z_I(t)$ is the transient position of the S/L interface with respect to the isotherm of the melting point of the pure solvent, $C_I(Z_I, t)$ is the transient solute concentration at the S/L interface, $D_L$ is the diffusion coefficient of solute in the liquid phase, $V_p$ and $G$ are the pulling velocity and temperature gradient in directional solidification, $k$ is the equilibrium partition coefficient for solute in solid phase, $m_L$ is the liquidus slope of the solution, $V_I = V_p + \frac{\partial Z_I(t)}{\partial t}$ is the transient velocity of the S/L interface with respect to the freezing platform. The initial conditions for the solidification system are

solute concentration in the liquid phase is uniform: $C_L(t=0) = C_\infty$ (Eq. 5)

the S/L interface position: $Z_I(t=0) = \Delta Z_1 = -\frac{m_L \cdot C_\infty}{G} = -\frac{\Delta T_f(C_\infty)}{G}$ (Eq. 6)

where $\Delta T_f(C_\infty)$ is the freezing point depression of solid phase at solute concentration $C_\infty$. By combining Eq. 1-4 via a simple boundary-layer approximation, the PDFs for description of transient S/L interface position $z_I(t)$ and time-dependent diffusion length $l(t)$ are proved to be [34]

$$\frac{\partial Z_I(t)}{\partial t} = \frac{2D_L \cdot (Z_I(t) - Z_\infty)}{l(t)(1-k)Z_I(t)} - V_p \tag{Eq. 7}$$

$$\frac{\partial l(t)}{\partial t} = \frac{4D_L \cdot (Z_I(t) - k \cdot Z_\infty)}{l(t)(1-k)Z_I(t)} - \frac{l(t)}{Z_I(t) - Z_\infty} \cdot \frac{\partial Z_I(t)}{\partial t} \tag{Eq. 8}$$

It is shown analytically that [34] for small times $t$

$$l(t) \approx \sqrt{\frac{8D_L \cdot t}{3}} \qquad \text{(Eq. 9)}$$

$$Z_I(t) = Z_\infty - V_p \cdot t + \sqrt{\frac{2D_L}{3} \cdot \frac{V_p}{|Z_\infty| \cdot (1-k)}} \cdot t^{3/2} \qquad \text{(Eq. 10)}$$

The equations are solved numerically with the following input parameters as shown in **Tab. 1**. By making the initial S/L interface position of sample solutions (i.e. the interface recoil $\Delta Z_1$ when $t=0$) of experiment and numerical solution via W-L model overlap, a comparison between experiments and predictions of W-L model will be made for both NaCl and PVA solutions as shown in the following section.

## 4. Results and discussions

**Figure 2** shows typical frames of in-situ movie which show the S/L interface positions for 0.1 M NaCl (up panel) and 10 wt.% PVA103 (down panel) solutions in comparison with that of pure water at some different time intervals of interest. The interface position as a function of time during transient process of planar instability is a direct indicator of solute redistribution in directional solidification [31]. **Figure 2** vividly demonstrates an example of how solute recoil ($\Delta Z_1$ and $\Delta Z_2$) was measured via DV method at different time intervals for both 0.1 M NaCl and 10 wt.% PVA103 solutions in our freezing stage. At $t=0$ s when samples were static, the S/L interface discrepancies between NaCl/PVA solutions and pure water were measured as $\Delta Z_1$. When $t > 0$ s, as time went by, the S/L interface discrepancies between NaCl/PVA solutions and pure water became larger and were measured as $\Delta Z_2$. And this measurement was conducted for both NaCl and PVA solutions with various solute concentrations as was mentioned in **Group A**. And the measured solute recoil $\Delta Z_2$ of **Group A** as a function of time were plotted in **Fig. 3** and **Fig. 4** for further analysis.

**Figure 3** shows $\Delta Z_2$ as a function of time for NaCl solutions with various initial concentrations (0.05M, 0.1M, 0.2M, 0.3M and 0.6M). It can be seen from **Fig. 3** that the experimental solute recoil of NaCl solutions was adequately described by the predictions of W-L model. The relative error of $\Delta Z_2$, defined as the absolute discrepancies between W-L model and experimental results divided by the predicted value via W-L model for NaCl solutions with various initial concentrations was also calculated and plotted in **Fig. 3** as was represented by the blue solid star. It can be seen in **Fig. 3** that the relative error was usually smaller than 10 %, which showed good agreement between experiments and W-L model. And the slight discrepancies may result from thickness effect in combination with gravity [38, 39], forming a S/L interface with "an asymmetrical wedged pattern" with a smaller thickness than the inner thickness of capillary, which reduced the solute pile-up compared to the case of a planar S/L interface before $t_{p-c}$. And when $t_{p-c}$ was reached, the experimental interface recoil $\Delta Z_2$ further deviated from the predictions of W-L model. In fact, W-L model is only valid for the prediction of the early stage of planar instability when instability does not occur and non-linear effect is negligible. And when perturbations develop, the system will enter non-linear regime.

**Figure 4** shows $\Delta Z_2$ as a function of time for PVA solutions with various initial concentrations and degree of hydrolysis (three initial concentrations of 0.25 wt.%, 5 wt.% and 10 wt.% and two degree of hydrolysis 96.8%-97.6% hydrolyzed and 86.5%-89% hydrolyzed). On the contrary, as shown in **Fig. 4**, for PVA solutions, there were severe discrepancies between the experimental data and the predictions of W-L model even at the very early stage of freezing. The possible reasons why W-L model failed to reproduce the experimental data for PVA solutions should be more complex than previous analysis on NaCl solutions. For PVA solutions with the lowest concentration (**Fig. 4 (a)**), the large discrepancies can be explained by the instability at the beginning of directional freezing. Since it has been verified both experimentally and computationally [40-43] that microscopic pits [40] would have formed due to the selective adsorption behavior of PVA macromolecules onto edge plane of ice, microscopically perturbed instead of planar S/L interface would have formed prior to directional freezing. Besides, solute may be easily engulfed into S/L interface due to the low diffusivity of PVA macromolecules, resulting less solute pile-up ahead of the S/L interface. Hence the experimental results of solute recoil were much smaller than predictions of W-L model from the beginning of directional freezing. For the discrepancies of PVA solutions with higher concentration (**Fig. 4 (b-d)**), severe discrepancies still existed. Besides the similar reasons for **Fig. 4 (a)**, concentration effect should also be taken into account. We examined the effect of non-linear concentration dependence of PVA diffusion coefficient $D_L$ on predicted results of W-L model since this dependence is common in polymer solution [18]. We compared a series of numerical results with varying $D_L$ ranging from 1.56 $um^2/s$ to 29 $um^2/s$ ( see **Supplementary Information A**). The comparison of solute recoil $\Delta Z_2$ of various $D_L$ was given in **Fig. A1**. From **Fig. A1** it was interesting to note that the decrease of $D_L$ by concentration effect can only enlarge the discrepancies instead of closer predictions to experimental results. Therefore, it is reasonable to suggest a different mechanism of solute recoil in directional freezing of PVA solutions. For PVA solutions, the ease of engulfment of PVA macromolecules into S/L interface and microscopic pits formed by PVA macromolecules prior to directional freezing should be responsible for the severe discrepancies in **Fig. 4**.

The unique solute recoil of PVA solution beyond the prediction of W-L model suggested a deviation from traditional diffusion-controlled planar instability. In order to further make a direct comparison of the planar instability between PVA and NaCl solutions, a side-by-side sample which consisted of one capillary with 0.5 wt.% NaCl solution and the other with 5wt.% PVA solution was unidirectionally solidified with different pulling velocities $V_p$ as was mentioned in **Group B**.

**Figure 5** shows some typical frames in the in-situ movie of planar instability morphology of 0.5 wt.% NaCl and 5wt.% PVA solutions under one given pulling velocity of $V_p$ = 20.34 um/s. And the planar instability morphology under other pulling velocities were included in **Supplementary Information B**. From **Fig. 5** and **Supplementary Information B**, it can be seen from an overall perspective that the plane-to-cell/dendrite transitions of 5 wt.% PVA103 solution (see left panel of **Fig. 5**) at different pulling velocities were always faster than 0.5 wt.% NaCl solution (see right panel of **Fig. 5**). And the faster pulling velocity would result in shorter duration after which the planar S/L interface become instable. It can also be seen from **Fig. 5** that there is a striking difference between NaCl solutions and PVA solutions in terms of their processes of planar instability.

For NaCl solutions (see right panel of **Fig. 5**), as the freezing process progressed, a solute concentration gradient was established in the vicinity of the freezing front. And a few cellular perturbations on the S/L interface occurred locally. And after an interval of 10s of seconds, the perturbations spread along the S/L interface and selection among perturbations occurred simultaneously to develop cellular (see **Fig. 5**)/dendritic (see **Supplementary Information B**) morphology. Distinct mushy zone among elongated cellular/dendritic arrays was clearly observed. This process was typical solute diffusion-controlled planar instability, which was commonly observed during solidification of alloy system [44]. On the contrary, for PVA solutions (see left panel of **Fig. 5**), spike-like perturbations seemed to develop in a global manner since perturbations simultaneously occurred across the whole S/L interface after a usually much shorter interval of freezing from static state. And mushy zone of PVA solution was hardly elongated, with only a small region near ice dendritic tips remaining unfrozen. And "ice-polymer composites" were formed which were composed of "unfrozen liquid microphase" rich in polymer coexists with ice platelets and gelation may occur afterwards [45, 46]. This fact revealed a very different nature of planar instability between NaCl solutions and PVA solutions morphologically. As have been suggested previously, a possible physical explanation is the selective adsorption behavior of PVA macromolecules onto edge plane of ice [40-43], resulting in the instability of the planar S/L interface with microscopic pits [40] formed prior to directional freezing.

Water permeation instead of diffusion of PVA in concentrated PVA solutions may also be germane to the abnormal solute recoil of concentrated PVA solutions in this paper. But the physics behind solvent permeation through random polymer network can be more sophisticated since one has to take into account many physical processes like formation of bounded water molecules via hydrogen-bond interaction with PVA macromolecules [45, 47-49] and other possible solute-solute/solvent interactions [50, 51]. To precisely reproduce the transient process of planar instability of polymer solution, a new model taking into account the peculiarity of polymer solution, though challenging, is to be established in the future.

## 5. Conclusion

In conclusion, the difference of planar instability of NaCl and PVA solutions was in-situ revealed both qualitatively and quantitatively in a unidirectional manner with controlled ice orientation. In the quantitative investigation, precise measurement of solute recoil $\Delta Z_2$ as a function of time has been made via DV method during the transient planar instability process for both NaCl and PVA solutions of various solute concentrations. It is found that W-L model can well reproduce the experimental results of NaCl solutions with different solute concentrations before planar instability occurs. On the contrary, for PVA solutions of all solute concentrations and degree of hydrolysis, there are severe discrepancies between experimental results and the predictions of W-L model. Such discrepancies may be resulted from complex interactions between PVA macromolecules and the S/L interface other than simple solute diffusion like ionic solution. In the in-situ comparison investigation, there is a striking difference between NaCl and PVA solutions in terms of their planar instability, with the one exhibiting a process from local instability to global instability and the other one exhibiting a global instability at the beginning of its planar instability, which is suggested to be a combined result of adsorption behavior of PVA macromolecules onto ice prior to freezing and the ease of engulfment of PVA macromolecules with low diffusivity into S/L interface. A physical picture concerning both possible S/L interface adsorption and water permeation-limited freezing in polymer solution is needed to better address the planar instability of polymer solutions in the future.

## Acknowledgements

This work was supported by the National Key R&D Program of China (Grant No.2018YFB1106003), National Natural Science Foundation of China (Grant No. 51701155), and the Fundamental Research Funds for the Central Universities (3102019ZD0402).

**Figure and table captions:**

**FIG. 1** The schematic diagram of experimental setup of customized horizontal Bridgeman freezing stage and measurement of S/L interface movement in combination with the differential visualization (DV) method via a side-by-side design of two parallel thin rectangular glass capillaries glued on a large silica glass sheet as freezing samples (solutions (up) and pure water (down)). Freezing samples are set within a one-dimensional imposed thermal gradient $\vec{G}$ established in a narrow gap between a cold copper block ($T_{cold}$) and a hot copper block ($T_{hot}$) connected to temperature controllers. When $t = 0$ s, freezing sample being static with zero pulling velocity, there is no thermal lag by instrumental recoil nor latent heat effect for the S/L interface of both pure water and solutions and only solute recoil of solutions is present, which is represented by $\Delta Z_1$; When $t > 0$ s, a certain pulling velocity $V_p$ is imposed to the sample and directional freezing occurs. The thermal lag by instrumental recoil and latent heat effect of each capillary is negligible compared to that of the glass sheet. The thermal lag of glass sheet is represented by $\Delta Z_3$. And solute recoil of a moving sample at velocity $V_p$ is represented by $\Delta Z_2$ which is a function of freezing time $t$. It should be noted that $\Delta Z_2$ is measured by CCD camera as the **solute recoil** via solute pile-up in the vicinity of the S/L interface during directional freezing in this paper. The crystal orientations of single crystal ice in each capillary for directional freezing are manipulated to be the same, satisfying the relation of direction $\vec{V}_p \parallel \vec{G} \parallel \{0001\}$ indicated by the solid arrow on the top of the figure.

**FIG. 2** Typical frames of in-situ movie of solute recoil of 0.1M NaCl (up panel) and 10 wt.% PVA103 (down panel) solutions as a function of time determined by DV method in the horizontal Bridgeman freezing stage of this work. The crystal orientations of single crystal ice in two capillaries for directional freezing were manipulated to be the same, satisfying the relation of direction $\vec{V}_p \parallel \vec{G} \parallel \{0001\}$ indicated by the solid arrow on the left of the figure. The scale bar in each figure was 250 um.

**FIG. 3** Solute recoil $\Delta Z_2$ of NaCl solutions with various initial concentrations as a function of time (black solid square) around the planar instability and comparison with the predicted results via W-L model (red solid circle). **(a)** 0.05 M NaCl with $V_p$ = 4.31 um/s and $G$ = 5.44 K/mm; **(b)** 0.1 M NaCl with $V_p$ = 4.26 um/s and $G$ = 5.39 K/mm; **(c)** 0.2 M NaCl with $V_p$ = 4.28 um/s and $G$ = 5.08 K/mm; **(d)** 0.3 M NaCl with $V_p$ = 4.24 um/s and $G$ = 5.90 K/mm; **(e)** 0.6 M NaCl with $V_p$ = 4.29 um/s and $G$ = 4.70 K/mm. The time intervals for observed plane-to-cell/dendrite transition for each figure ($t_{p-c}$) were indicated.

**FIG. 4** Solute recoil $\Delta Z_2$ of PVA solutions with various initial concentrations as a function of time (black solid square) around the planar instability and comparison with the predicted results via W-L model (red solid circle). **(a)** 0.25 wt.% PVA103 with $V_p$ = 5.77 um/s and $G$ = 5.44 K/mm; **(b)** 5 wt.% PVA203 with $V_p$ = 4.15 um/s and $G$ = 4.86 K/mm; **(c)** 5 wt.% PVA103 with $V_p$ = 3.86 um/s and $G$ = 5.44 K/mm; **(d)** 10 wt.% PVA103 with $V_p$ = 3.98 um/s and $G$ = 5.51 K/mm. The time intervals for observed plane-to-cell/dendrite transition ($t_{p-c}$) for each figure except **(a)** were indicated.

**FIG. 5** Typical frames of in-situ comparison of planar instability of S/L interfaces for 5 wt.% PVA103 solution (left panel) and 0.5 wt.% NaCl solution (right panel) under fixed pulling velocities $V_p$ = 20.34 um/s and thermal gradient $G$ = 3.18 K/mm. The crystal orientations of single crystal ice in two capillaries for directional freezing were manipulated to be the same, satisfying the relation of direction $\vec{V}_p \parallel \vec{G} \parallel \{0001\}$ indicated by the solid arrow on the left of each figure. The scale bar in each figure was 250 um.

**Tab. 1** The parameters used in the numerical solution of W-L model with a time step $\Delta t$ of 0.1 s. The equilibrium partition coefficient $k$ is taken as 0 due to extremely

low solubility of solute molecules into ice lattice [52, 53]. The freezing point depression $\Delta T_f$ was precisely determined by cooling curve method (see **Supplementary Information C**). The diffusion coefficients $D_L$ for 5 wt.% PVA103 and 0.5 wt.% NaCl are taken as 15.6 $um^2/s$ and 774 $um^2/s$, respectively (see the determination for the used values of $D_L$ in **Supplementary Information D**).

**Supplementary Information A: Comparison between experimental results and the numerical results via W-L model with varying diffusion coefficients of 10 wt.% PVA103 solution;**

**Supplementary Information B: In-situ comparison of planar instability of S/L interfaces for 5 wt.% PVA103 solution and 0.5 wt.% NaCl solution under different pulling velocities;**

**Supplementary Information C: Measured freezing point depression of NaCl solutions and PVA203/PVA103 solutions;**

**Supplementary Information D: Diffusion coefficients of 5 wt.% PVA103 and 0.5 wt.% NaCl aqueous solutions used in W-L model of this paper.**

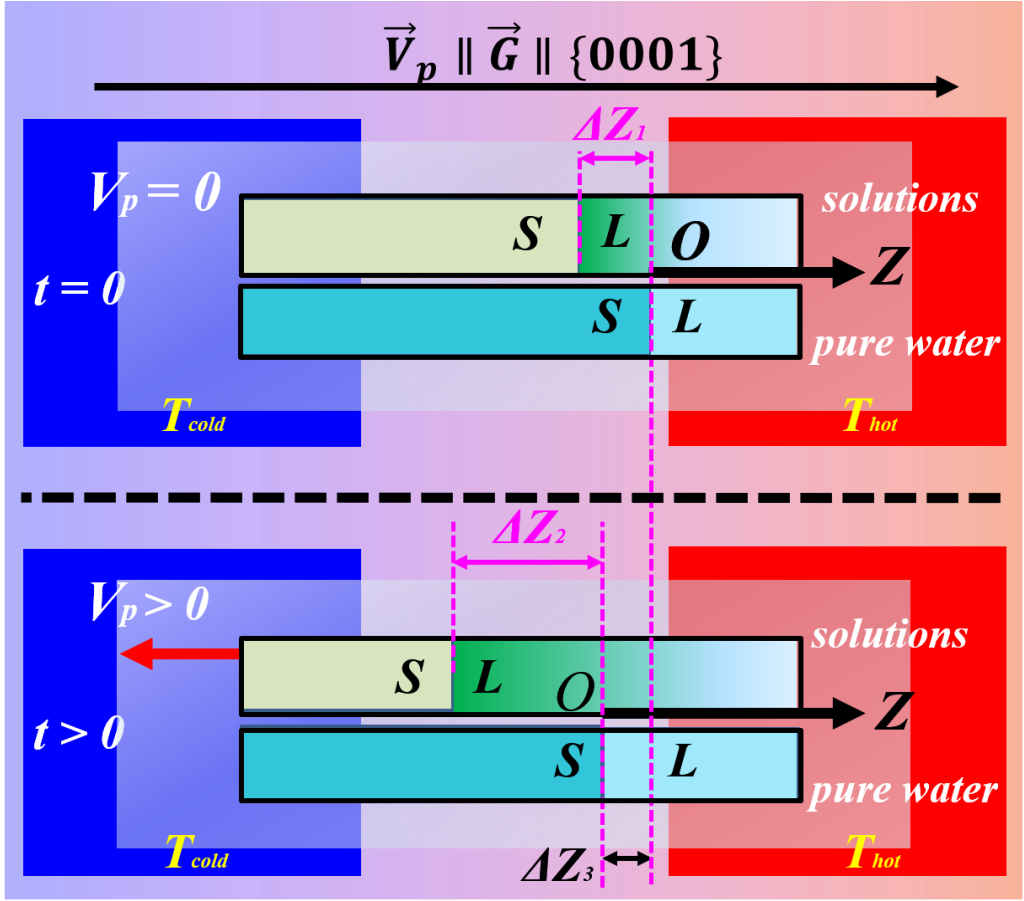

**FIG. 1** The schematic diagram of experimental setup of customized horizontal Bridgeman freezing stage and measurement of S/L interface movement in combination with the differential visualization (DV) method via a side-by-side design of two parallel thin rectangular glass capillaries glued on a large silica glass sheet as freezing samples (solutions (up) and pure water (down)). Freezing samples are set within a one-dimensional imposed thermal gradient $\vec{G}$ established in a narrow gap between a cold copper block ($T_{cold}$) and a hot copper block ($T_{hot}$) connected to temperature controllers. When $t = 0$ s, freezing sample being static with zero pulling velocity, there is no thermal lag by instrumental recoil nor latent heat effect for the S/L interface of both pure water and solutions and only solute recoil of solutions is present, which is represented by $\Delta Z_1$; When $t > 0$ s, a certain pulling velocity $V_p$ is imposed to the sample and directional freezing occurs. The thermal lag by instrumental recoil and latent heat effect of each capillary is negligible compared to that of the glass sheet. The thermal lag of glass sheet is represented by $\Delta Z_3$. And solute recoil of a moving sample at velocity $V_p$ is represented by $\Delta Z_2$ which is a function of freezing time $t$. It should be noted that $\Delta Z_2$ is measured by CCD camera as the **solute recoil** via solute pile-up in the vicinity of the S/L interface during directional freezing in this paper. The crystal orientations of single crystal ice in each capillary for directional freezing are manipulated to be the same, satisfying the relation of direction $\vec{V}_P \parallel \vec{G} \parallel \{0001\}$ indicated by the solid arrow on the top of the figure.

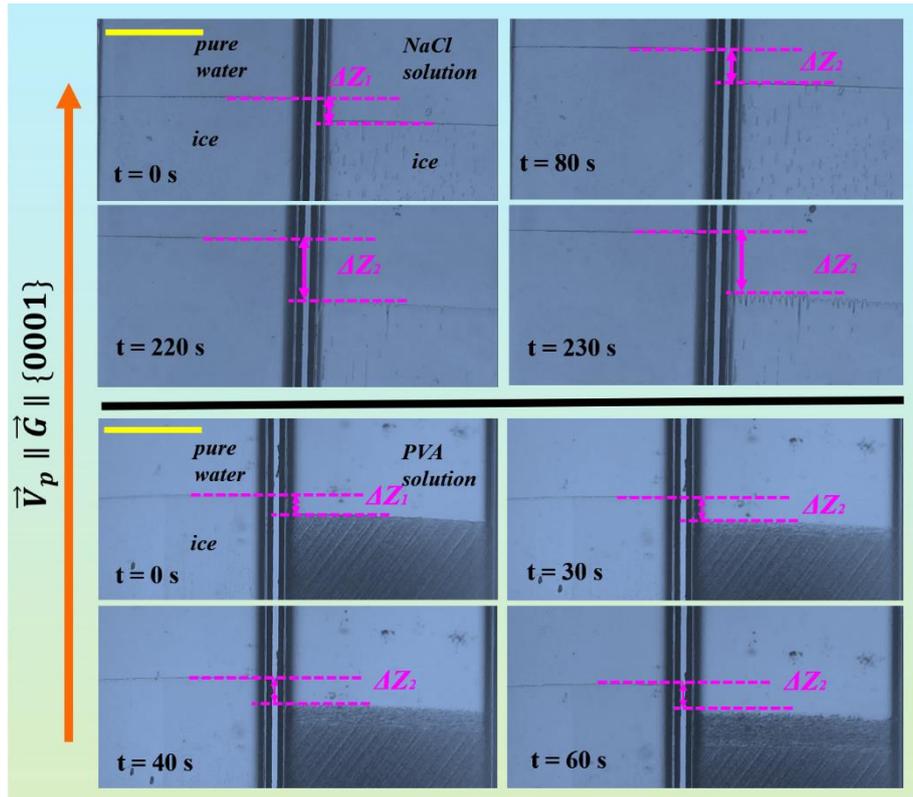

**FIG. 2** Typical frames of in-situ movie of solute recoil of 0.1M NaCl (up panel) and 10 wt.% PVA103 (down panel) solutions as a function of time determined by DV method in the horizontal Bridgeman freezing stage of this work. The crystal orientations of single crystal ice in two capillaries for directional freezing were manipulated to be the same, satisfying the relation of direction $\vec{V}_p \parallel \vec{G} \parallel \{0001\}$ indicated by the solid arrow on the left of the figure. The scale bar in each figure was 250 um.

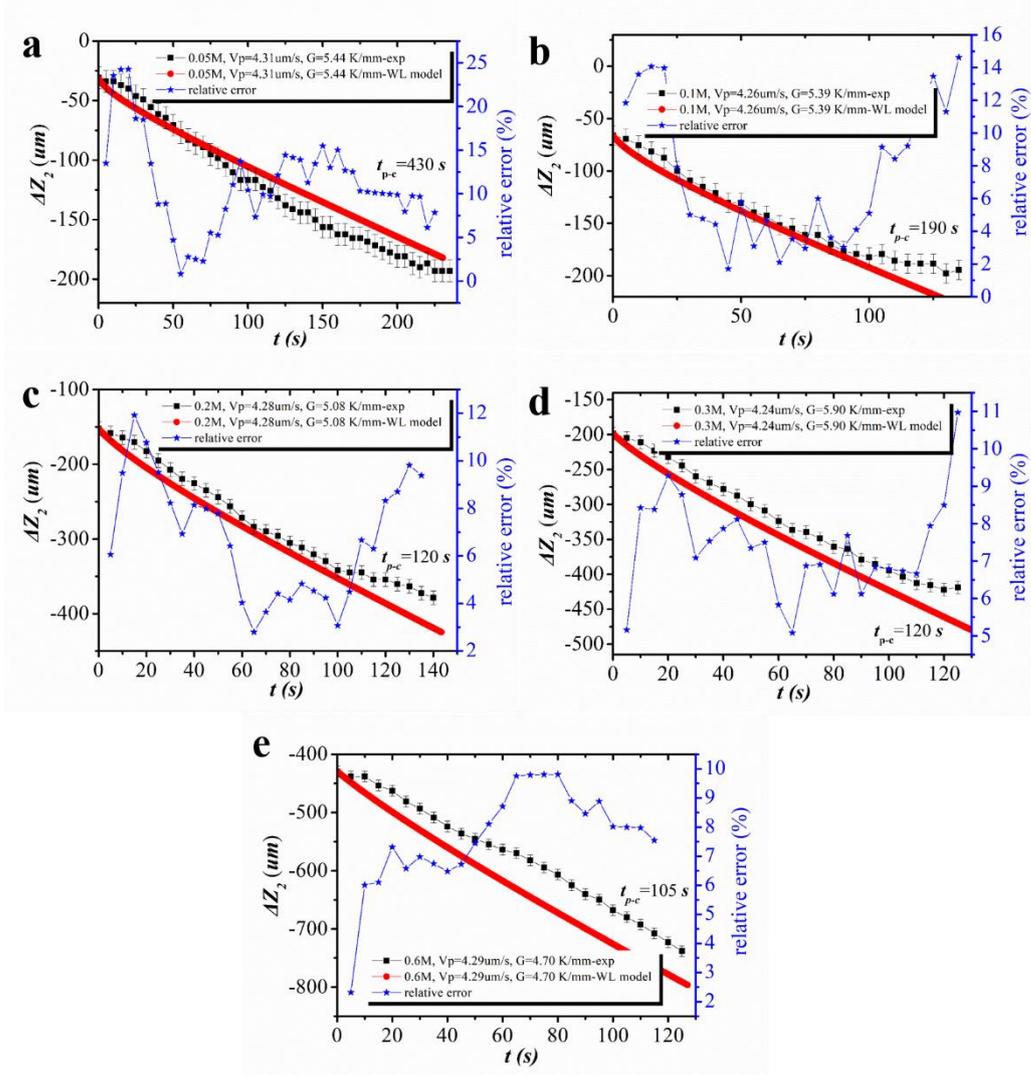

**FIG. 3** Solute recoil $\Delta Z_2$ of NaCl solutions with various initial concentrations as a function of time (black solid square) around the planar instability and comparison with the predicted results via W-L model (red solid circle) and the corresponding relative error (blue solid star). **(a)** 0.05 M NaCl with $V_p$ = 4.31 um/s and $G$ = 5.44 K/mm; **(b)** 0.1 M NaCl with $V_p$ = 4.26 um/s and $G$ = 5.39 K/mm; **(c)** 0.2 M NaCl with $V_p$ = 4.28 um/s and $G$ = 5.08 K/mm; **(d)** 0.3 M NaCl with $V_p$ = 4.24 um/s and $G$ = 5.90 K/mm; **(e)** 0.6 M NaCl with $V_p$ = 4.29 um/s and $G$ = 4.70 K/mm. The time intervals for observed plane-to-cell/dendrite transition for each figure ($t_{p-c}$) were indicated.

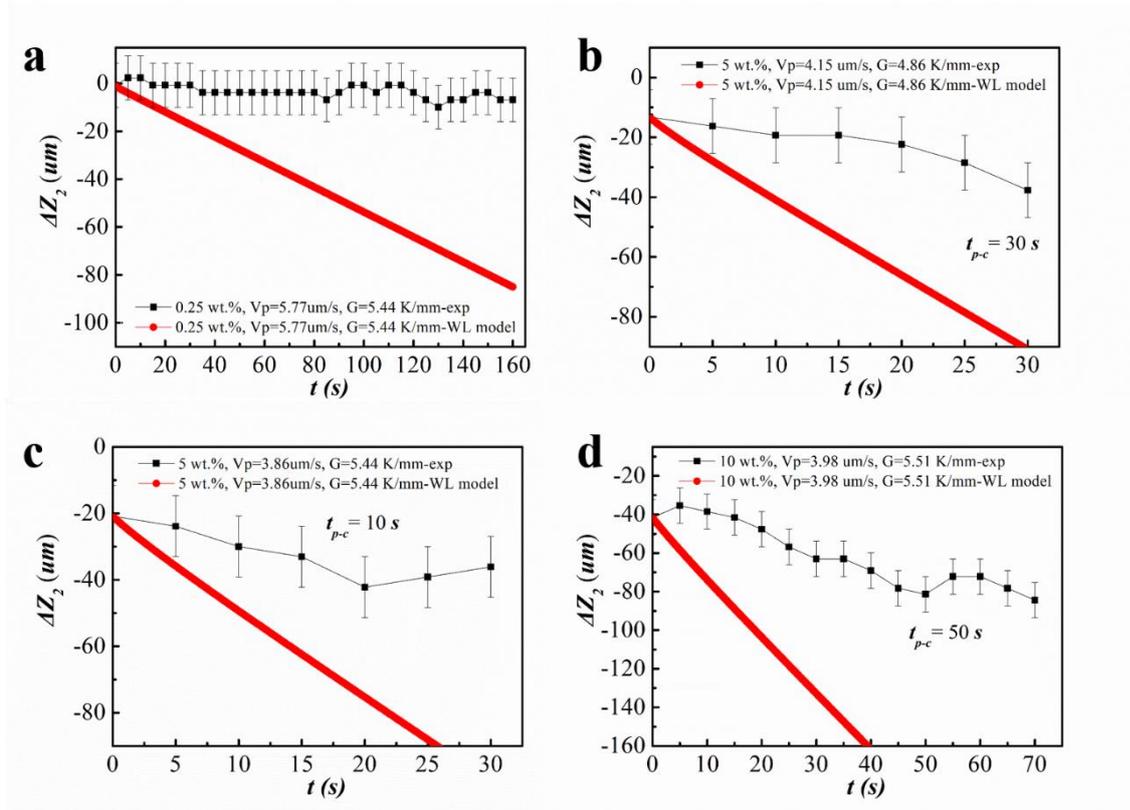

**FIG. 4** Solute recoil $\Delta Z_2$ of PVA solutions with various initial concentrations as a function of time (black solid square) around the planar instability and comparison with the predicted results via W-L model (red solid circle). **(a)** 0.25 wt.% PVA103 with $V_p$ = 5.77 um/s and $G$ = 5.44 K/mm; **(b)** 5 wt.% PVA203 with $V_p$ = 4.15 um/s and $G$ = 4.86 K/mm; **(c)** 5 wt.% PVA103 with $V_p$ = 3.86 um/s and $G$ = 5.44 K/mm; **(d)** 10 wt.% PVA103 with $V_p$ = 3.98 um/s and $G$ = 5.51 K/mm. The time intervals for observed plane-to-cell/dendrite transition ($t_{p-c}$) for each figure except **(a)** were indicated.

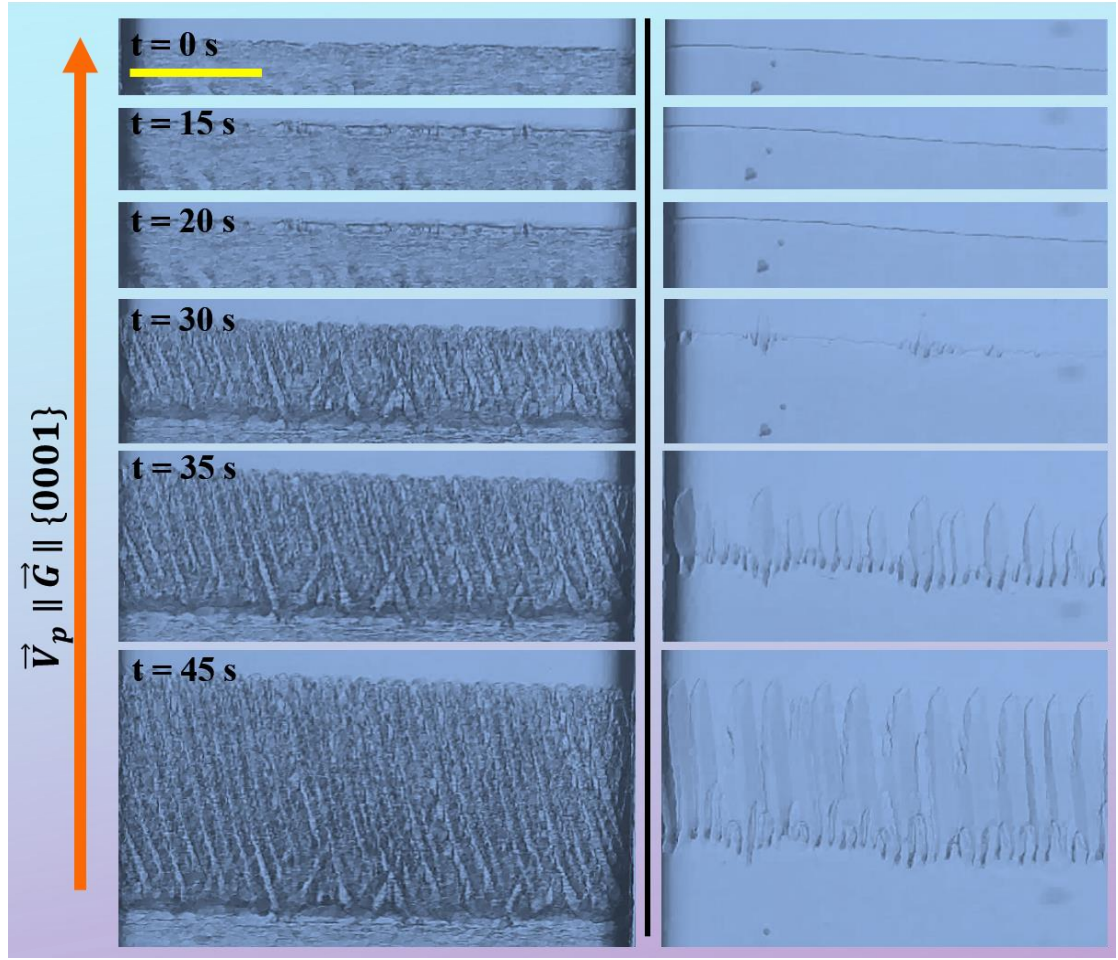

**FIG. 5** Typical frames of in-situ comparison of planar instability of S/L interfaces for 5 wt.% PVA103 solution (left panel) and 0.5 wt.% NaCl solution (right panel) under fixed pulling velocities $V_p$ = 20.34 um/s and thermal gradient $G$ = 3.18 K/mm. The crystal orientations of single crystal ice in two capillaries for directional freezing were manipulated to be the same, satisfying the relation of direction $\vec{V}_P \parallel \vec{G} \parallel \{0001\}$ indicated by the solid arrow on the left of each figure. The scale bar in each figure was 250 um.

**Tab. 1** The parameters used in the Finite Difference method with a time step $\Delta t$ of 0.1 s. The equilibrium partition coefficient $k$ is taken as 0 due to extremely low solubility of solute molecules into ice lattice [52, 53]. The freezing point depression $\Delta T_f$ was precisely determined by cooling curve method (see **Supplementary Information C**). The liquidus slopes $m_L$ of both NaCl solution (within the whole range of concentration) and PVA solution (concentration range of up to 10 wt.%) were determined by the data in **Supplementary Information C**. The diffusion coefficients $D_L$ for 5 wt.% PVA103 and 0.5 wt.% NaCl are taken as 15.6 $um^2/s$ and 774 $um^2/s$, respectively (see the determination for the used values of $D_L$ in **Supplementary Information D**).

| Solute | $m_L$(K/wt.%) | $C_\infty$ | $V_p$ (um/s) | $G$ (K/mm) | $\Delta T_f$(K) | $\Delta Z_1$ (um) |
|---|---|---|---|---|---|---|
| NaCl | 0.592 | 0.05 M | 4.31 | 5.44 | 0.168 | -30.882 |
| | | 0.1 M | 4.26 | 5.39 | 0.356 | -66.048 |
| | | 0.2 M | 4.26 | 5.08 | 0.773 | -152.165 |
| | | 0.3 M | 4.24 | 5.90 | 1.172 | -198.644 |
| | | 0.6 M | 4.29 | 4.70 | 2.017 | -429.149 |
| PVA103 | 0.0236 | 0.25wt.% | 3.95 | 5.77 | 0.004 | -0.6932 |
| | | 5 wt.% | 3.86 | 4.76 | 0.099 | -20.7983 |
| | | 10 wt.% | 3.86 | 5.51 | 0.229 | -54.332 |
| PVA203 | 0.0194 | 5 wt.% | 4.15 | 4.86 | 0.064 | -13.169 |